\documentclass[a4paper, amsfonts, amssymb, amsmath, reprint, showkeys, nofootinbib, twoside,superscriptaddress]{revtex4-2}
\usepackage[english]{babel}
\usepackage[utf8]{inputenc}
\usepackage[colorinlistoftodos, color=green!40, prependcaption]{todonotes}
\usepackage{hyperref}
\usepackage[caption=false]{subfig}
\usepackage{physics}
\usepackage{mathrsfs}
\usepackage{float}
\usepackage{subfig}
\usepackage{graphicx,epstopdf}
\begin{document}
\setlength{\parskip}{0pt}
\title{Dissipative phase transitions in optomechanical systems}
\author{Fatemeh Bibak}
\affiliation{University of Vienna, Faculty of Physics, Vienna Center for Quantum Science and Technology,
 Boltzmanngasse 5, 1090 Vienna, Austria
}
\affiliation{Institute for Quantum Optics and Quantum Information (IQOQI) Vienna,
Austrian Academy of Sciences,
Boltzmanngasse 3, A-1090 Vienna, Austria}
\author{Uro\v s Deli\'c}
\affiliation{University of Vienna, Faculty of Physics, Vienna Center for Quantum Science and Technology,
 Boltzmanngasse 5, 1090 Vienna, Austria
}
\author{Markus Aspelmeyer}
\affiliation{University of Vienna, Faculty of Physics, Vienna Center for Quantum Science and Technology,
 Boltzmanngasse 5, 1090 Vienna, Austria
}
\affiliation{Institute for Quantum Optics and Quantum Information (IQOQI) Vienna,
Austrian Academy of Sciences,
Boltzmanngasse 3, A-1090 Vienna, Austria}
\author{Borivoje Daki\'c}
\affiliation{University of Vienna, Faculty of Physics, Vienna Center for Quantum Science and Technology,
 Boltzmanngasse 5, 1090 Vienna, Austria
}
\affiliation{Institute for Quantum Optics and Quantum Information (IQOQI) Vienna,
Austrian Academy of Sciences,
Boltzmanngasse 3, A-1090 Vienna, Austria}
    %\email[Correspondence email address: ]{email@institution.com}% Your name
\date{\today} % Leave empty to omit a date

\begin{abstract}
We show that optomechanical quantum systems can undergo dissipative phase transitions within the limit of small nonlinear interaction and strong external drive. In such a defined thermodynamical limit, the nonlinear interaction stabilizes optomechanical dynamics in strong and ultrastrong coupling regimes. As a consequence optomechanical systems possess a rich phase diagram consisting of periodic orbits, discontinuous, and continuous dissipative phase transitions with and without bifurcation. We also find a critical point where continuous and discontinuous dissipative phase transition lines meet. Our analysis demonstrates that optomechanical systems are valuable for understanding the rich physics of dissipative phase transitions and ultrastrong coupling regimes.
%, where observation of these effects is possible even with a finite number of components, essentially due to their infinite-dimensional Hilbert spaces.
\end{abstract}
\maketitle
\section{Introduction}
One of the recent research foci in the field of open quantum systems are so-called dissipative phase transitions (DPTs)~\cite{spectraltheory}. They are defined as an effect of abrupt change in the physical properties of the steady-state caused by small variations of the system external parameters in the appropriate thermodynamic limit. Unlike their equilibrium counterparts, which can be divided into two distinct categories based on the nature of the fluctuations (thermal and quantum phase transitions), DPTs generally support the coexistence of thermal and quantum fluctuations~\cite{dpt10}, which leads in general to richer phase diagrams. DPTs are typically accompanied by intriguing phenomena~\cite{dpt1,dpt2,dpt3,dpt4,dpt5,dpt6,dpt7,dpt8,dpt9,Centralspin,kerr1,kerr2}, such as  critical slowing down~\cite{criticalslowingdown}, optical bistability~\cite{ob}, and break down of photon blockade~\cite{jaynes}. Some of these effects have been  experimentally studied in a variety of physical systems~\cite{exp1,exp2,exp3,exp4,exp5,Cai21}.

In the context of optical quantum systems, recent developments in control and manipulation, as well as their interaction with the environment, make them good candiates for simulating the physics of many-body systems, and quantum phase transitions~\cite{m1,m2,m3,m4,m5,m6,m7,osborn}. A fundamental distinction between phase transitions in quantum optical systems and conventional phase transitions arises from the notion of the thermodynamic limit. The latter is usually associated with the limit of a large number of particles: only in this regime can one observe a phase transition. While phase transitions in quantum optical systems can fall into this category (e.g., the Dicke model~\cite{Dicke,closeDicke}), the observation of phase transitions is possible even with a finite number of components due to their infinite-dimensional Hilbert spaces~\cite{Rabi,Hwang16}. In this letter, we follow this line of research and show that two coupled quantum harmonic oscillators exhibit a very rich phase diagram encompassing all known types of DPTs. A physical system exemplified in detail is the dissipative optomechanical system driven by an external laser field~\cite{OM}. The interplay between the weak nonlinear interaction and a strong external drive (in the exact thermodynamical limit) results in linear dynamics with various steady states associated with different phase-space regions. The stability diagram identifies these regions and their separation is marked by transition lines at which dissipative phase transitions occur. This is in contrast to a simplified standard linearized optomechanical approach, where the system dynamics is considered to be unstable beyond the transition lines ~\cite{vitali,peterson,hofer}. Here we demonstrate that stabilization of the system is possible through the dissipative phase transitions in the whole red- and partially in the blue-detuned regime. Our analysis reveals a rich phase diagram composed of discontinuous DPTs, continuous DPTs, and periodic orbits. The continuous DPT appears in two distinct categories: with and without bifurcation. We find a single critical point where bifurcation occurs in complete analogy to the continuous phase transition with symmetry breaking. The continuous DPT without bifurcation (symmetry breaking) is a curious effect that has only very recently been theoretically studied in non-equilibrium quantum systems~\cite{withoutbif}. To confirm the quantum properties of DPTs, we provide a numerical study and show that quantum entanglement and squeezing are maximized along transition lines, thus correctly marking the phase transition. Finally, we specify the universality class of the optomechanical systems by studying finite-size scaling and critical exponents.

\section{Quantum Optomechanics in the thermodynamic limit}
\textit{\textbf{Thermodynamic limit.}}
A generic quantum optomechanical system consists of a laser-driven optical cavity with a movable mirror. The Hamiltonian of the system is typically of the following form~\cite{cavityom} (with $\hbar=1$):
 \begin{align}
     H =\,&\omega_c\, a^\dagger a+\omega_m\, b^\dagger b+ g_0\, a^\dagger a(b+b^\dagger)\\\nonumber&-i\,(E^*e^{i\omega_Lt}a-E e^{-i\omega_Lt}a^\dagger).
 \end{align}
Here, $a$ is the annihilation operator of the cavity mode with frequency $\omega_c$, and $b$ is the annihilation operator of the mechanical resonator mode with frequency $\omega_m$. The coupling rate between the cavity field and the mechanical resonator is denoted by $g_0$, and $E$ is the amplitude of the laser drive. Without loss of generality, we assume $E$ is a real number. In the frame rotating with the laser frequency $\omega_L$, the Hamiltonian becomes time-independent:
\begin{equation}
     H =-\Delta\, a^\dagger a+\omega_m\, b^\dagger b+ g_0\, a^\dagger a(b+b^\dagger)-i\,E\,(a-a^\dagger),
 \end{equation}
where $\Delta=\omega_L-\omega_c$ denotes detuning. The Markovian open system is described by a time-independent master equation
\begin{equation}
     \label{master}\Dot{\rho} =\mathcal{L}[\rho]=-i[H,\rho]+\kappa\,\mathcal{D}[a]\rho+\gamma\,\mathcal{D}[b]\rho,
 \end{equation}
 where $\kappa$ and $\gamma$ are the damping rates of the cavity and the mechanical resonator, respectively. Here $\mathcal{D}[o]\rho=2o\rho o^\dagger-\rho o^\dagger o-o^\dagger o\rho$ is the Lindblad operator.

We begin our analysis by changing the picture $\rho\mapsto U(t)\rho U^{\dagger}(t)$, with
\begin{equation}
U(t)=\exp[\alpha^*(t) a-\alpha(t) a^\dagger] \exp[\beta^*(t) b-\beta(t) b^\dagger]
\end{equation}
being the displacement operator and $\alpha(t)$ and $\beta(t)$ are yet to be determined. The master equation \eqref{master} keeps the form in the new picture with the new Linbland $\mathcal{L'}[\rho]=-i[H(t),\rho]+\kappa\,\mathcal{D}[a]\rho+\gamma\,\mathcal{D}[b]\rho$ and the new time-dependent Hamiltonian which we split into three terms
\begin{equation}\label{Ham}
  H(t)=H^{{(1)}}(t)+H^{{(2)}}(t)+H_{int}.
\end{equation}
%\begin{widetext}
These are
\begin{eqnarray}\label{H1(t)}
 H^{(1)}(t)&=&i\Big[\Dot{\alpha^*}+\left(i\Delta-ig_0(\beta+\beta^*)+\kappa\right)\alpha^*-E\Big]a\\\nonumber
 &&+i\left[\Dot{\beta^*}-(i\omega_m-\gamma)\beta^*-ig_0\abs{\alpha}^2\right]b+h.c.,\\\label{H2(t)}
 H^{(2)}(t)&=&-\left(\Delta-g_0\beta-g _0\beta^*\right)a^\dagger a+\omega_m b^\dagger b\\\nonumber
 &&+g_0\left(\alpha^* a+\alpha a^\dagger\right)(b+b^\dagger),\\\label{H3(t)}
 H_{int}&=&g_0a^\dagger a(b+b^\dagger).
\end{eqnarray}
%\end{widetext}
Here we omit the explicit time dependence $\alpha=\alpha(t)$ and $\beta=\beta(t)$ for simplicity. Our goal is to show that only the quadratic term $H^{(2)}(t)$ survives in the thermodynamic limit, leading to an exact linear dynamics of the system. The necessary precondition is that the linear term vanishes, i.e., $H^{(1)}(t)=0$, which gives the set of equations for $\alpha(t)$ and $\beta(t)$:
\begin{eqnarray}\label{scl}
     \Dot{\alpha}(t)&=&(i\,\Delta-i\,g_0(\beta(t)+\beta^*(t))-\kappa)\,\alpha(t)+E,\\
     \Dot{\beta}(t)&=&-(i\,\omega_m+\gamma)\,\beta(t)-i\,g_0\abs{\alpha(t)}^2.
 \end{eqnarray}
These are the semiclassical equations of motion. We define the thermodynamic limit as follows
\begin{equation}\label{TDL}\tag{{\bf{TDL}}}
  E\rightarrow\infty~\mathrm{and}~g_0\rightarrow0,~~\mathrm{s.t.}~~ \Tilde{E}=g_0E\rightarrow\mathrm{const.}
\end{equation}
%To support this choice, consider a statistical mechanics analogy, e.g., gas with $n$ particles confined in a volume $V$. The standard thermodynamic limit is reached for $n\rightarrow\infty$ and $V\rightarrow \infty$, keeping the density $N/V$ constant. In our system, one can take the number of photons $n\sim E$ and identify $1/V\sim g_0$. The reason for the latter condition is as follows.
Given the first condition $E\rightarrow\infty$, the solution to the semiclassical equations \eqref{scl} are not well-defined, since in this case we get $\alpha(t)\rightarrow\infty$ and $\beta(t)\rightarrow\infty$. However, multiplying \eqref{scl} with the interaction constant $g_0$ from both sides, we obtain
\begin{align}\label{TDscl}
     &\Dot{\tilde{\alpha}}(t)=\left(i\,\Delta-i\,\left(\tilde{\beta}(t)+\tilde{\beta}^*(t)\right)-\kappa\right)\,\tilde{\alpha}(t)+\Tilde{E},\nonumber\\
     &\Dot{\tilde{\beta}}(t)=-(i\,\omega_m+\gamma)\,\tilde{\beta}(t)-i\,\abs{\tilde{\alpha}(t)}^2,
 %    \label{SemiCL}
 \end{align}
where $\tilde{\alpha}(t)=g_0\alpha(t)$ and $\tilde{\beta}(t)=g_0\beta(t)$ represent the re-scaled mean amplitudes of the cavity field and the mechanical resonator of the steady-state. This is directly measurable in the experiment~\cite{Uros} and used by default in theoretical analysis. These equations now have well defined solutions in the limit $E\to+\infty$ for $\tilde{\alpha}(t)$ and $\tilde{\beta}(t)$ only if $g_0\to0$ and $\Tilde{E}=g_0E\rightarrow\mathrm{const.}$ This choice of \eqref{TDL} is also in analogy with statistical mechanics, e.g., the TDL for a gas with $n$ particles confined in a volume $V$ is reached as $n\rightarrow\infty$ and $V\rightarrow \infty$, keeping the density $N/V$ constant. Following this analogy in optomechanical systems, we have the mean photon number in the cavity  $n\propto E$ and the length of the cavity $L\propto 1/g_0$~\cite{OM}, which is then consistent with the choice of the \eqref{TDL} .

Once we set the \eqref{TDL} condition, the displaced nonlinear interaction term \eqref{H3(t)} $H_{int}=g_0a^\dagger a(b+b^\dagger)\rightarrow0$ vanishes as $g_0\to0$, while the only term that survives in the limit is the quadratic Hamiltonian \eqref{H2(t)} as a function of $\tilde{\alpha}(t)$ and $\tilde{\beta}(t)$. Thus, the dynamics becomes linear in the thermodynamical limit.

To study the steady-state properties, we take the limit $t\rightarrow+\infty$ for which equations \eqref{TDscl} can lead to stationary solutions $(\tilde{\alpha}_s,\tilde{\beta}_s)$ obtained by solving $\Dot{\tilde{\alpha}}(t)=0$ and $\Dot{\tilde{\beta}}(t)=0$, i.e.
 \begin{align}\label{SSE}
     &\left(i\,\Delta-i\,\left(\tilde{\beta}_s+\tilde{\beta}^*_s\right)-\kappa\right)\,\tilde{\alpha}_s+\Tilde{E}=0,\nonumber\\
     &-(i\,\omega_m+\gamma)\,\tilde{\beta_s}-i\,\abs{\tilde{\alpha_s}}^2=0.
 %    \label{SemiCL}
 \end{align}
 In this case, the Hamiltonian in \eqref{Ham}(and consequently the corresponding master equation) becomes time-independent.

\textbf{\textit{Semi-classical solutions.}}\label{stb} Before proceeding to the quantum regime, we shall find the stationary solutions of the semiclassical equations \eqref{SSE}. These are the set of coupled cubic equations which can be reduced to the second-order using the following ansatz. First, we define the rescaled mean photon number $\tilde{n}=\abs{\tilde{\alpha}}^2$, which satisfies the following equation:
 \begin{align}\label{mean1}
     &[(\Delta+\frac{2 \tilde{n}\omega_m}{\omega_m^2+\gamma^2})^2+\kappa^2]\tilde{n}-\tilde{E}\,^2=0.
 \end{align}
In general, there are three solutions to the cubic equation above; however, only those satisfying $\tilde{n}\geq0$ are physical. It is well known that one solution, which we label as $\tilde{n}_1$, is always physical in the whole parameter space~\cite{OM}. We set $\tilde{n}_1$ as the reference and the corresponding solutions of \eqref{SSE} we denote as $(\tilde{\alpha_1},\tilde{\beta_1})$, i.e., those satisfying $\tilde{n}_1=|\tilde{\alpha}_1|^2$. On this basis, we define two key parameters, effective detuning, and coupling strength:
\begin{eqnarray}
\tilde{\Delta}&=&\Delta-(\tilde{\beta}_1+\tilde{\beta}_1^*),\\\nonumber
G&=&\sqrt{\tilde{n}_1}=\abs{\tilde{\alpha}_1}.
\end{eqnarray}
Having assumed that one solution of the cubic equation \eqref{mean1} is known (parameter $\tilde{n}_1=G^2$), we can find the analytical expression for the other two solutions
 \begin{eqnarray}
    \tilde{n}_{2,3}&=&\frac{1}{2}G^2-\frac{\tilde{\Delta}}{2\omega_m}(\gamma^2+\omega_m^2)\\
    &&\pm\frac{1}{2}\sqrt{G^4-\frac{\kappa^2}{\omega_m^2}(\gamma^2+\omega_m^2)^2-2G^2\frac{\tilde{\Delta}}{\omega_m}(\gamma^2+\omega_m^2)}.\nonumber
 \end{eqnarray}
Finally, the stationary solutions to the semiclassical equations \eqref{SSE} can be expressed solely in terms of parameters $\tilde{\Delta}$ and $G$, i.e.
 \begin{eqnarray}
     \tilde{\alpha}_j&=&\frac{i G\sqrt{\tilde{\Delta}^2+\kappa^2}}{\tilde{\Delta}+\frac{2\omega_m (\tilde{n}_j-\tilde{n}_1)}{\omega_m^2+\gamma^2}+i\kappa},\\
     \tilde{\beta}_j&=&\frac{-\abs{\tilde{\alpha}_j}^2}{\omega_m-i\gamma},~~j=1,2,3.\nonumber
 \end{eqnarray}
Each of these solutions, when substituted into \eqref{H2(t)}, defines one Hamiltonian in the thermodynamical limit, i.e.
  \begin{equation}\label{TDHam}
   H_j=-(\Delta-\tilde{\beta}_j-\tilde{\beta}_j^*)a^\dagger a+\omega_m b^\dagger b%\\\nonumber
   +\left(\tilde{\alpha}_j^* a+\tilde{\alpha}_j a^\dagger\right)(b+b^\dagger),
 \end{equation}
with $j=1,2,3$. This is our working Hamiltonian in the thermodynamic limit, which has two new properties: a) the detuning of the cavity mode is shifted to the value $\tilde{\beta}_j+\tilde{\beta}_j^*$, and b) the interaction between the modes is linear and its strength is directly proportional to the amplitude $|\tilde{\alpha}_j|$. Moreover, these Hamiltonians depend only on parameters $\tilde{\Delta}$ and $G$, which will serve to define our stability diagram. For example, the Hamiltonian for the reference solution reads
\begin{equation}
    H_1=-\tilde{\Delta}a^\dagger a+\omega_m b^\dagger b+G(e^{i\theta(\tilde{\Delta})}a+e^{-i\theta(\tilde{\Delta})}a^\dagger)(b+b^\dagger),
\end{equation}
with $\theta=\mathrm{Arg}[\tilde{\alpha}_1]=\mathrm{Arg}[\kappa+i \tilde{\Delta}]$.

 \textbf{\textit{Stability analysis and quantum steady-state.}}\label{fokker}
In this section, we examine the quantum mechanical behavior of the system in the thermodynamic limit. To do this, we find the steady-state solution of the master equation
\begin{equation}\label{TDmaster}
     \Dot{\rho} =\mathcal{L}[\rho]=-i[H_j,\rho]+\kappa\,\mathcal{D}[a]\rho+\gamma\,\mathcal{D}[b]\rho,~~j=1,2,3.
 \end{equation}
For simplicity, we use an equivalent formulation of the problem in terms of the Fokker-Planck equation for the Wigner quasi-probability distribution associated with the density operator $\rho$ \cite{Mandel}. Firstly, we introduce the cavity field quadratures, $ x_c\equiv( a+ a^\dagger)/\sqrt{2}$ and $p_c\equiv ( a- a^\dagger)/i \sqrt{2}$, and the mechanical resonator quadratures, $ x_m\equiv(b+ b^\dagger)/\sqrt{2}$, and $ p_m\equiv ( b- b^\dagger)/i \sqrt{2}$. The Fokker-Planck equation in terms of quadrature vector in phase space $\Vec{x}=(x_c, p_c, x_m, p_m)$ can be stated as follows:
\begin{align}\label{Fokker}
    \partial_t W(\Vec{x},t)=(-\Vec{\nabla} A_j\Vec{x}+\frac{1}{2}\Vec{\nabla}^T D\Vec{\nabla}) \,W(\Vec{x},t),
\end{align}
with diffusion matrix $D=diag(\kappa,\kappa,\gamma,\gamma)$ and $A_j$ being the drift matrix which can be calculated directly from the Hamiltonian \eqref{TDHam}, i.e.
\begin{small}
\begin{align}
    &A_j=\\\nonumber&\begin{pmatrix}-\kappa&{\scriptstyle -2(\frac{\omega G^2}{\omega^2+\gamma^2}+  \mathrm{Re}[\tilde{\beta}_j])+\tilde{\Delta}}&{\scriptstyle 2\mathrm{Im}[\tilde{\alpha}_j]}&0\\{\scriptstyle2(\frac{\omega G^2}{\omega^2+\gamma^2}+\mathrm{Re}[\tilde{\beta}_j])-\tilde{\Delta}}&-\kappa&{\scriptstyle 2\mathrm{Re}[\tilde{\alpha}_j]}&0\\0&0&-\gamma&-\omega_m\\{\scriptstyle 2\mathrm{Re}[\tilde{\alpha}_j]}&{\scriptstyle -2\mathrm{Im}[\tilde{\alpha}_j]}&\omega_m&-\gamma\end{pmatrix}.
\end{align} 
\end{small}

The drift matrix $A_j$ controls the time evolution of the first moment  $\langle x\rangle(t)$
\begin{align}
&\label{firstmom}\frac{d\langle\Vec{x}\rangle}{dt}=A_j\langle\Vec{x}\rangle,
\end{align}
while the time evolution of the second moment, correlation matrix $V$, is given by
\begin{align}
    &\label{secondmom}\frac{dV}{dt}=A_jV+VA_j^T+D.%\nonumber
\end{align}
Since the dynamics is linear, the steady-state solution to the Eq. (\ref{Fokker}) is a unique Gaussian state (provided that the dynamics is stable):
\begin{align}\label{tgaussian}
    W_{s}(\Vec{x})=\frac{1}{4\pi^2\sqrt{\mathrm{det} V_s}} e^{\frac{1}{2}(\Vec{x}- \Vec{x_s})^TV_s^{-1}(\Vec{x}- \Vec{x_s})}.
\end{align}
Here $x_s$ and $V_s$ are stationary solution to Eq. (\ref{firstmom}) and Eq. (\ref{secondmom}), respectively.

Each $A_j$ (for $j=1,2,3$) defines one steady-state solution which we label with roman numerals I (reference), II and III. We study their stability by considering the time evolution of a small deviation around the stationary solution, $x = x_s+\delta x$,
\begin{align}
    \frac{d\langle \delta x\rangle}{dt}=A_j\langle \delta x\rangle,
\end{align}
which has the time evolution $\langle \delta x\rangle(t)=e^{t A_j}\langle\delta x\rangle(0)$. The steady-state is stable if all eigenvalues $\lambda_i$ of $A_j$ satisfy $\mathrm{Re}[\lambda_i]<0$ (more details about eigenvalues of the drift matrix can be found in Appendix~\ref{eigs}). Using this condition we obtain the necessary and sufficient conditions for the stability of the solution, known as the Routh-Hurwitz criterion~\cite{Routh}:
\begin{align}
     &\label{hard}4\gamma\kappa[(\gamma+\kappa)^2+(2\frac{\omega_m G^2}{\omega_m^2+\gamma^2}+\tilde{\beta}_i+\tilde{\beta}^*_i+\omega_m-\tilde{\Delta})^2]\nonumber\\\,\,\,&[(\gamma+\kappa)^2+(-2\frac{\omega_m G^2}{\omega_m^2+\gamma^2}-\tilde{\beta}_i-\tilde{\beta}^*_i+\omega_m+\tilde{\Delta})^2]\nonumber\\\,\,\,&+16(\gamma+\kappa)^2\omega_m\abs{\tilde{\alpha}_i}^2(2\frac{\omega_m G^2}{\omega_m^2+\gamma^2}+\tilde{\beta}_i+\tilde{\beta}_i^*-\tilde{\Delta})> 0,\\
    &\label{soft}(\gamma^2+\omega_m^2)(\kappa^2+(2\frac{\omega_m G^2}{\omega_m^2+\gamma^2}+\tilde{\beta}_i+\tilde{\beta}_i^*-\tilde{\Delta})^2)\nonumber\\&\,\,\,-4\omega_m\abs{\tilde{\alpha}_i}^2(2\frac{\omega_m G^2}{\omega_m^2+\gamma^2}+\tilde{\beta}_i+\tilde{\beta}_i^*-\tilde{\Delta})> 0.
 \end{align}
Instability occurs if at least one of the above inequalities is violated. This is closely related to the dissipative phase transition, which we will examine in the next section.

\section{Dissipative phase transitions}
When an abrupt change in the physical properties of the steady-state occurs (in the thermodynamic limit), we have a dissipative phase transition. Mathematically, this situation happens when the so-called Liouvilian gap closes~\cite{spectraltheory}. More precisely, consider the solution of the eigenvalue problem of the operator in \eqref{TDmaster}
\begin{equation}
\mathcal{L}[e_k]=\mu_k e_k,~~~ k=1,2,...
\end{equation}
Given its specific form one can show that all the eigenvalues satisfy $\mathrm{Re}[\mu_k]\leq0$, with $\mu_1=0$ being the eigenvalue associated to the steady-state~\cite{57,58}. For convenience, we sort the eigenvalues in such a way that $|\mathrm{Re}[\mu_1]|<|\mathrm{Re}[\mu_2]|<...$, and the relevant quantity is the Liouvilian gap $\mu=-\mu_2$. The DPT occurs when the real part of the gap vanishes, i.e., $\mathrm{Re}[\mu]=0$. For linear systems, this condition completely translates to a constraint on eigenvalues of the drift matrix~\cite{gap,Dicke}, i.e., for eigenvalues of $A_j$ sorted in order $|\mathrm{Re}[\lambda_1]|<|\mathrm{Re}[\lambda_2]|<|\mathrm{Re}[\lambda_3]|<|\mathrm{Re}[\lambda_4]|$, the DPT occurs for $\mathrm{Re}[\lambda_1]=0$. The last condition is precisely the Routh-Hurwitz criterion \eqref{hard} and \eqref{soft} defined in the previous section. In this way, we establish a one-to-one correspondence between instability and dissipative phase transitions for our optomechanical system. The points for which the left hand side of Eq. \eqref{hard} vanishes are called hard-mode instabilities and correspond to the pure imaginary Lindblad gap $\mathrm{Re}[\lambda_1]=0$. In this regime the system exhibits \emph{periodic orbits}~\cite{periodicorbit}. On the other hand, the points for which the left hand side of Eq. \eqref{soft} vanishes, are called soft-mode instabilities and correspond to the closure of the Lindblad gap, i.e. $\lambda_1=0$. In this regime the system undergoes \emph{first}- and \emph{second-order} dissipative phase transitions~\cite{spectraltheory}.
\begin{figure}
       \includegraphics[width=0.8\linewidth]{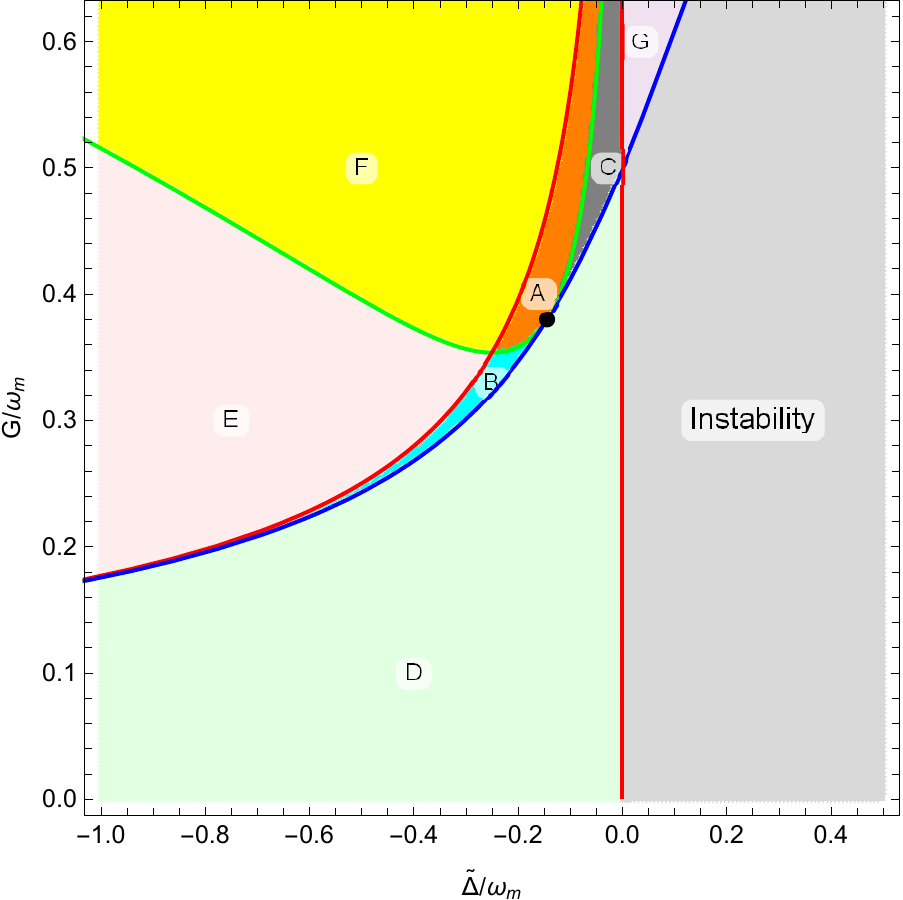}
       \captionsetup{justification=justified,singlelinecheck=false}
    \caption{\textbf{Phase (stability) diagram.} Optomechanical systems possess three distinct steady-state solutions which we label as I (reference), II and III. Based on the stability regions of these solutions, the stability diagram can be decomposed into seven regions ($A$ to $G$). Two stable steady states coexist in regions $A$ (II and III), $B$ (I and III), and $C$ (I and II). Only one stable steady-state exist in regions $D$ (I), $E$ (I), $F$ (II), and $G$ (II). The solutions are separated by instability lines: the hard mode (red) and the soft mode (blue and green) with one touching point (black dot, critical point). For simplicity, the results are evaluated for $\kappa=1/4$ and $\gamma=0$ in units $\omega_m=1$. For other values the results do not change qualitatively.}
    \label{fig:softhard}
\end{figure}

\begin{figure}[!ht]
\subfloat[]{\includegraphics[width=4.2cm]{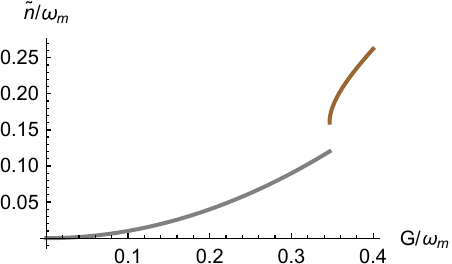}}
        \subfloat[]{\includegraphics[width=4.2cm]{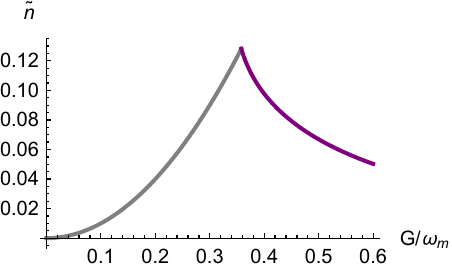}}
    \captionsetup{justification=justified,singlelinecheck=false}
    \caption{\textbf{Rescaled mean photon number $\tilde{n}$.} a) Discontinuous DPT between the I and the III steady-state. Continuous DPT between the I and the II steady-state. b)  The plots are evaluated for $\tilde{\Delta}=-0.2$, $\kappa=1/4$, and $\gamma=0$ in units $\omega_m=1$.  }
    \label{fig:mean}
\end{figure}

We investigate the stability of all three solutions, and the results are shown in the phase (stability) diagram in Fig. \ref{fig:softhard}. The diagram is composed of seven regions (described in detail in the caption of Fig.\ref{fig:softhard}) separated by hard- (red lines) and soft-mode (blue and green lines) instability lines. Interestingly, we find stability in the ultrastrong coupling regime (typically $G>0.1~\omega_m$), in the whole red- and partially in the blue-detuned regimes, where the stabilization is obtained via dissipative phase transition. This contrasts with the common belief that this regime has fundamental parametric instability (see Appendix~\ref{eigs}). DPTs occur at transition lines, and we divide them into several distinct categories:%We identify seven different regions separated by the instability lines at which the DPT occurs. They fall into several distinct categories:

\textbf{\textit{First-order phase transition.}} First-order phase transitions or discontinuous phase transitions correspond to a discontinuous change in the behavior of an observable's mean values across the transition line. The typical behavior is shown in Fig.~\ref{fig:mean}~(a), where the rescaled mean photon number shows a discontinuous change. The optomechanical system exhibits a first-order phase transition along the blue-soft mode instability line (see Fig.~\ref{fig:softhard}). For $\tilde{\Delta}_c<\tilde{\Delta}<0$ (with $\tilde{\Delta}_c/\omega_m=-1/4\sqrt{3}$) we have a first-order DPT between the steady states I and II, whereas, for $\tilde{\Delta}<\tilde{\Delta}_c$ we have a first-order DPT between the steady states I and III. Notably, the latter occurs in the regime of parameters which is in the domain of experimental reach \cite{peterson}. The first-order transition line separates bistable regions, i.e., B and C, from the monostable region D. Typically, the system's state in these bistable regions depends on its history, which then gives rise to hysteresis behavior \cite{OM,Ghobadi}.  

\begin{figure}[!ht]
  \centering
       \includegraphics[width=5cm]{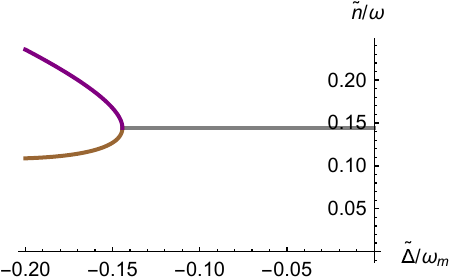}
    \caption{\textbf{Bifurcation diagram.} The bifurcation in the rescaled mean photon number in the cavity at the critical point, $(\tilde{\Delta}_c,G_c)=(-1/4\sqrt{3},\sqrt{1/4\sqrt{3}})$. These plots are evaluated for $G=G_c$, $\kappa=1/4$, and $\gamma=0$ in units $\omega_m=1$.}
    \label{fig:bifurcation}
\end{figure}

\textbf{\textit{Second-order phase transition.}} Second-order or continuous phase transitions involve a continuous change in an observable's mean values across the transition line. The typical behavior is shown in Fig.~\ref{fig:mean}(b), where the rescaled mean photon number displays a continuous change across the phase transition. %Continuous DPTs happen when the soft mode instability lines of at least two different semi-classical steady states in the thermodynamics limit are equal .
In our system, the second-order DPT occurs along the soft-mode instability line where two different steady-states in the thermodynamic limit are equal. We find two categories of continuous DPTs: with and without bifurcation. DPTs without bifurcation occur when the soft-mode instability of exactly two steady states are equal such that each of the solutions is stable on one side of the stability line. In our case, such phase transition between the steady states I and III occurs along the green soft mode instability line for $\tilde{\Delta}_c<\tilde{\Delta}<0$ (see Fig. \ref{fig:softhard}). Similarly, in the region defined by $\tilde{\Delta}<\tilde{\Delta}_c$ along the green instability line, we have another continuous DPT without bifurcation between the steady states I and II. The two soft-mode instability lines (blue and green) meet at the critical point $(\tilde{\Delta}_c, G_c)$ where we have a continuous DPT with bifurcation. This behavior is depicted in Fig. \ref{fig:bifurcation}. 

\begin{figure}[!ht]
  \centering
       \subfloat[]{\includegraphics[width=4.2cm]{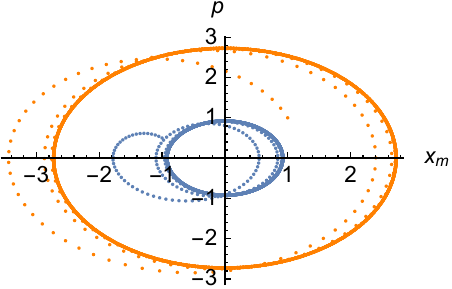}}
         \subfloat[]{\includegraphics[width=4.2cm]{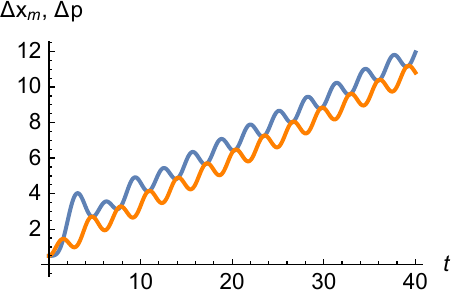}}
    \caption{\textbf{Periodic orbits.} The optomechanical systems exhibit periodic behavior along hard-mode instability line (red lines in Fig.~\ref{fig:softhard}). Panel (a) depicts the mechanical phase space. The system dynamic is sensitive to the initial condition. Blue and orange orbits correspond to following initial conditions $(x_c,p_c,x_m,p_m)=(1,1,-1,1/2)$, $(1,1/2,1,1)$ respectively. Panel (b) displays fluctuations in mechanical resonator position (blue) and momentum (orange) growing linearly in time, indicating the instability of periodic orbits along the hard mode instability line. The plots are calculated from the solutions of Eq.~\eqref{secondmom} for $V(0)=\mathbb{I}$. These graphs have been calculated for $\kappa=1/4$, $\gamma=0$, $\tilde{\Delta}=-1$, and $G=1/(4\sqrt{2})$ in units $\omega_m=1$.}
    \label{fig:periodicorbit}
\end{figure}

\textbf{\textit{Periodic orbits.}} The imaginary gap of the Liouvillian gives rise to the onset of oscillations, which do not have any counterpart in the closed system. These oscillations can be divided into limit cycles and periodic orbits~\cite{periodicorbit}. Limit cycles are isolated trajectories, meaning all neighboring trajectories either converge to the limit cycle or diverge to another attractor. Stable limit cycles are accompanied by self-sustained oscillations and have been extensively studied in optomechanical systems \cite{LC1,LC2,LC3,LC4}. In contrast, periodic orbits are surrounded by closed orbits and thus the amplitude of oscillations depends on the initial conditions. Optomechanical systems experience periodic orbits along the hard-mode instability lines. The typical behavior is shown in Fig.~\ref{fig:periodicorbit}. 

\textit{\textbf{Quantum properties.}} An interesting question is the behavior of quantum properties such as entanglement and squeezing in different phases of the system. To answer this question, one must consider quantum fluctuations. In the thermodynamic limit, stable linear dynamics results in a Gaussian steady-state, which is entirely characterized by its correlation matrix that is found as the stationary solution of Eq. (\ref{secondmom}).

\begin{figure}[!ht]
  \centering
       \subfloat[]{\includegraphics[width=4.2cm]{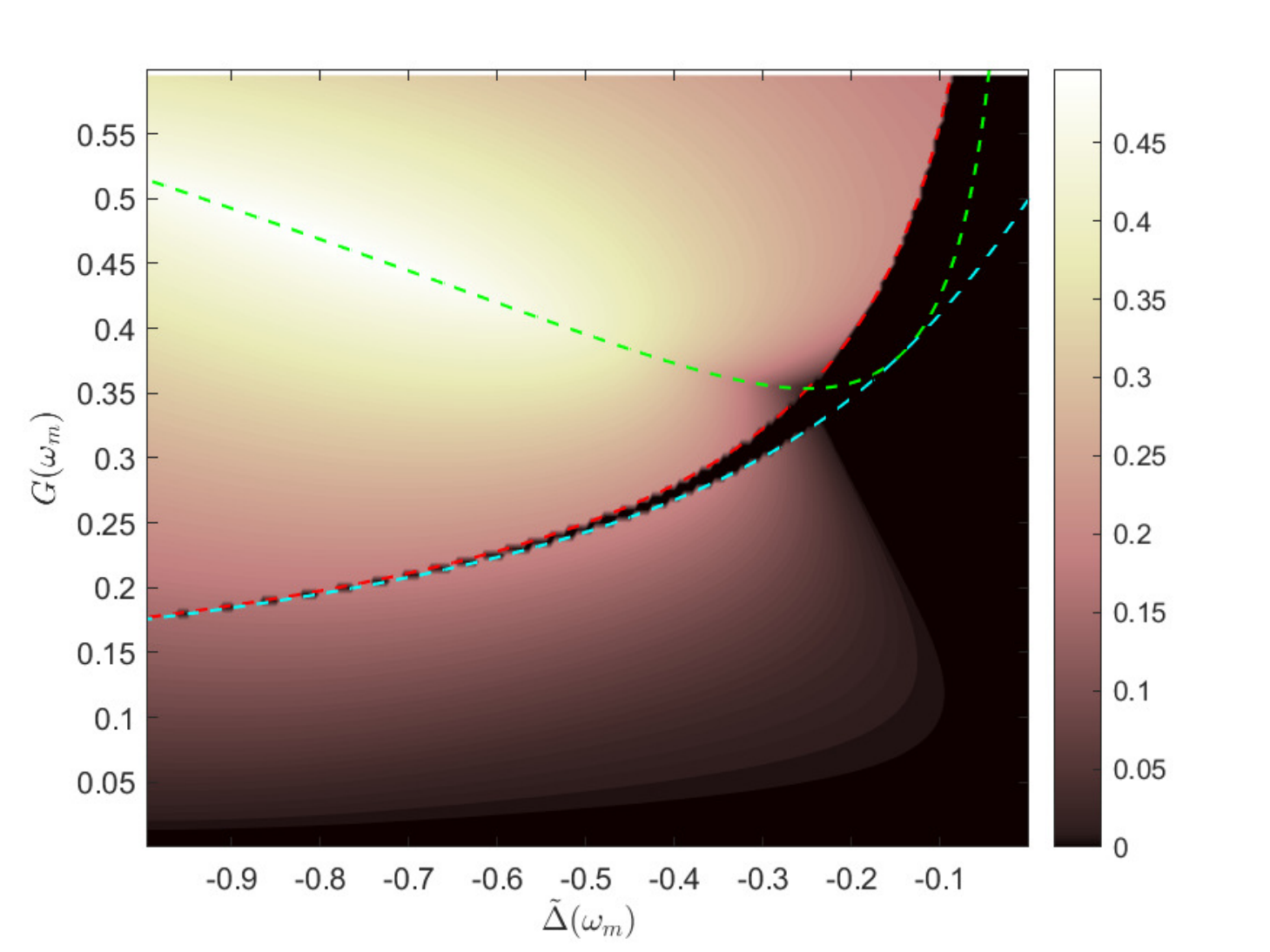}}
         \subfloat[]{\includegraphics[width=4.2cm]{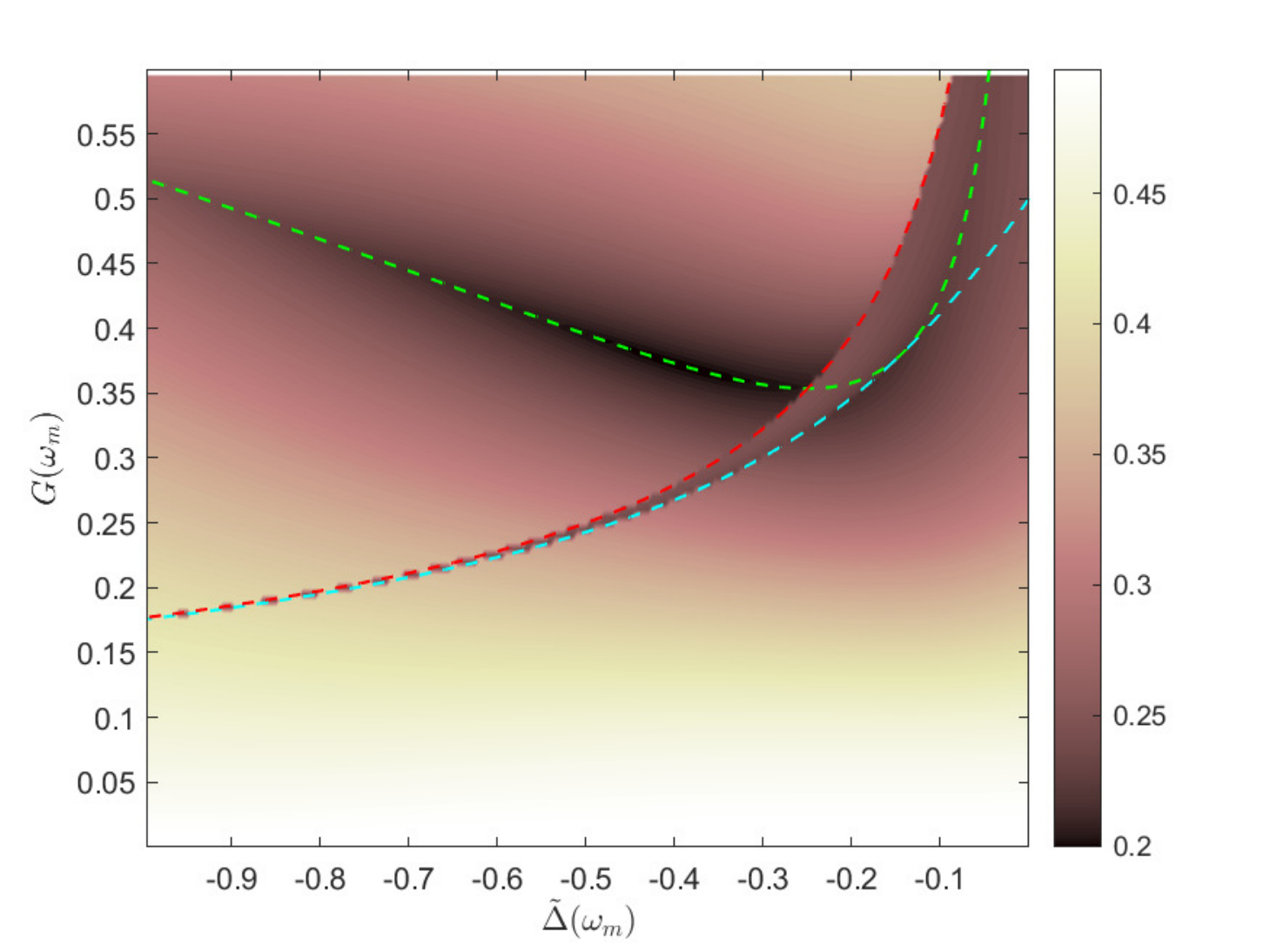}}
    \caption{\textbf{Entanglement and squeezing}. a) Logarithmic negativity in the phase diagram. b) The squeezing in system as measured via the smallest eigenvalue of the correlation matrix in the phase diagram. In the above diagrams, the considered steady states in regions with more than one stable steady-state are as follows (regions are defined in Fig.\ref{fig:softhard}): $A$ (III), $B$ (III), $C$ (I). Here $\kappa=1/4$, and $\gamma=0$ in units $\omega_m=1$.  }
    \label{fig:sqzent}
\end{figure}

To study entanglement between the mechanical and cavity modes we use Simon's criteria~\cite{simon}. We employ logarithmic negativity as a measure of entanglement for continuous variable systems which is defined as $E_\mathcal{N}=\max[0,-\ln 2\eta^-]$~\cite{logneg2}. For a two-mode Gaussian state with a correlation matrix of the form
\begin{align}
   V=\begin{pmatrix}
    \alpha&\beta\\\beta^T&\gamma\
\end{pmatrix},
\end{align}
we have $\eta^-\equiv2^{-1/2}[\Sigma(V)-\sqrt{\Sigma(V)^2-4 \mathrm{det} V}]^{1/2}$, where $\Sigma(V)\equiv \mathrm{det}\, \alpha+\mathrm{det}\, \gamma-2 \mathrm{det} \,\beta$. A Gaussian state is entangled if and only if $\eta^-<1/2$. The typical behavior of the logarithmic negativity is shown in Fig.\ref{fig:sqzent} (a). Entanglement reaches a maximum value of $ E_\mathcal{N}=1/2$ at the continuous DPT between steady states I and II, thus correctly marking the dissipative phase transition in the system.

Next, we perform squeezing analysis in the phase diagram. A Gaussian state is called a squeezed state if there exists a basis in phase space in which at least one diagonal element of the correlation matrix is smaller than 1/2 (shot-noise limit)~\cite{sqz}. We evaluate the diagonal elements of the correlation matrix in the $(x_c,p_c,x_m,p_m)$ basis, and we find all of them to be greater than 1/2 in the whole phase space. This means that measurement of quadratures will not display squeezing effects at any point in the space of parameters. However, investigation of the lowest eigenvalue of the correlation matrix, presented in Fig.~\ref{fig:sqzent} (b), reveals that there exist regions with eigenvalues less than $1/2$. This means that squeezing is present for a hybrid mode which involves combination of the cavity and mechanical mode~\footnote{The hybrid modes can be obtained by applying appropriate linear optical elements (e.g., beam splitters and phase shifters) to the cavity and mechanical modes.}. As can be seen from the evaluated plot, the maximum squeezing also occurs at the continuous DPT line between steady states I and II. These calculations show that squeezing properly marks the phase transition. While squeezing changes continuously along continuous DPTs, it changes discontinuously along discontinuous DPTs. This observation shows that squeezing also reveals the nature of the DPTs.

\begin{figure}[!t]
  \centering
       \subfloat[]{\includegraphics[width=4.42cm]{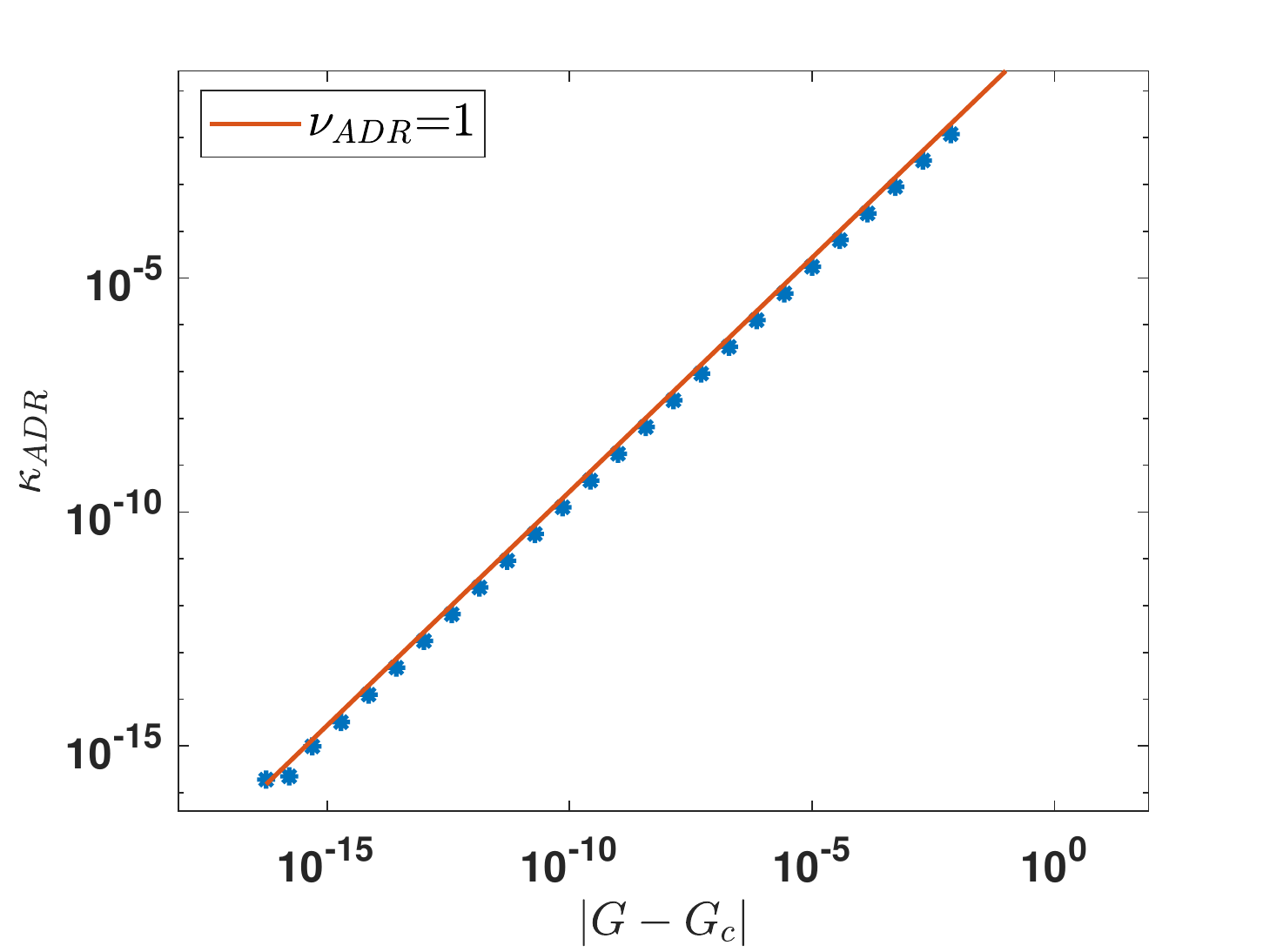}}
     \subfloat[]{\includegraphics[width=4.42cm]{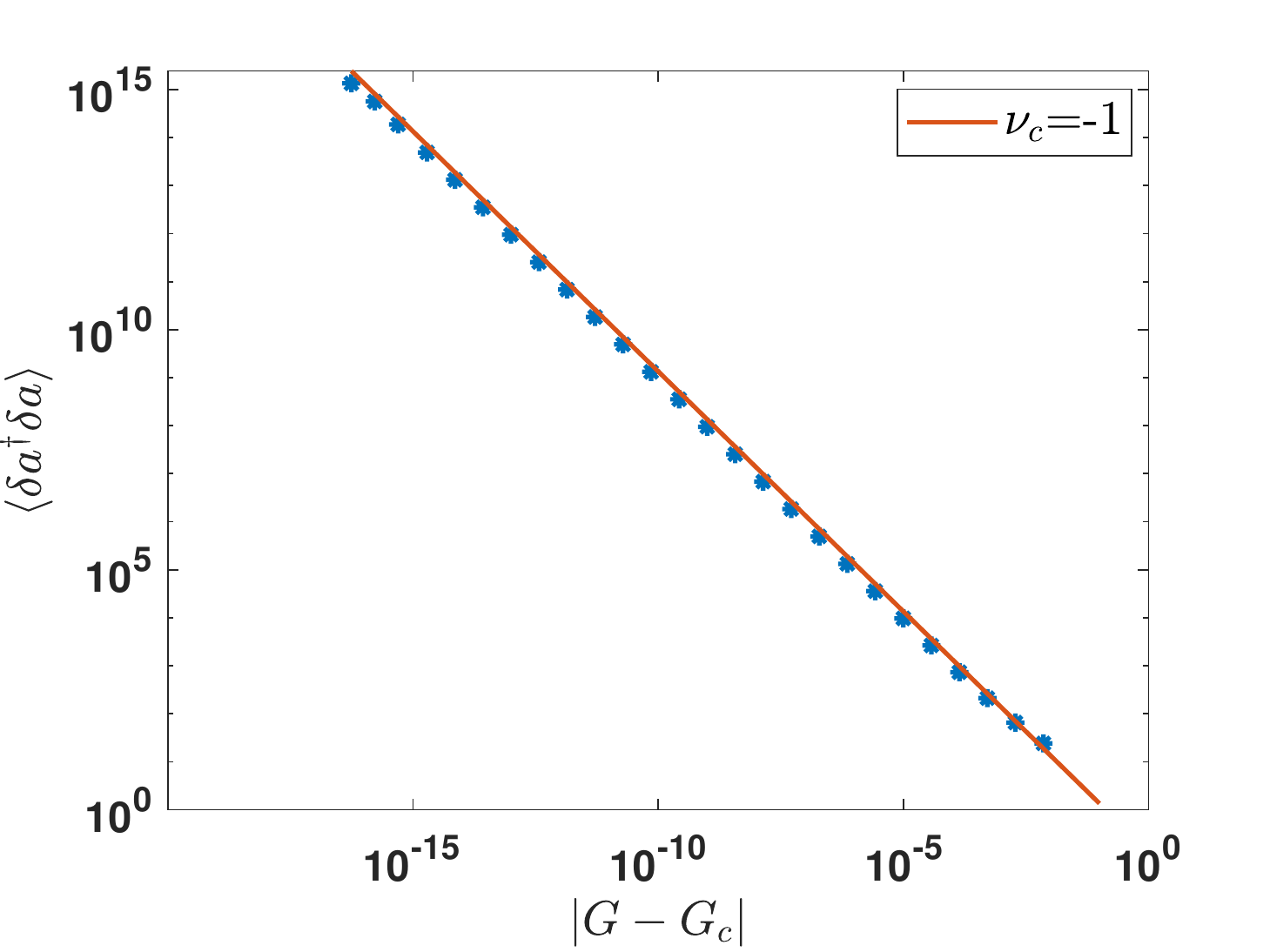}}
    \caption{\textbf{The critical exponents.} a) Asymptotic decay rate, $\kappa_{ADR}$ in terms of $\abs{G-G_c}$ in the log-log plot displaying powerlaw behavior near the continuous DPT point with exponent $\nu_{ADR}=1$. b) Mean fluctuations of photon number in the cavity $\langle\delta a^\dagger \delta a\rangle$ in terms of $\abs{G-G_c}$ in the log-log plot exhibiting powerlaw behavior near the continuous DPT point with exponent $\nu_{c}=-1$. Both plots have been evaluated for $\kappa=1/4$, $\gamma=0$, and $\tilde{\Delta}=\tilde{\Delta}_c$ in units $\omega_m=1$. }
    \label{fig:powerlaw1}
\end{figure}

\textbf{\textit{Universality Class}}. Universality is one of the remarkable properties of continuous equilibrium phase transitions~\cite{RG} which originates from the long-range fluctuation close to the continuous phase transition point. In such cases, the correlation length becomes much larger than the typical range of the interactions, which results in the cooperative phenomena independent of the microscopic details of the considered system. Unlike for equilibrium phase transitions, little is known about non-equilibrium phase transitions due to the lack of a general framework for studying DPTs. Still, it is believed that non-equilibrium phenomena can be grouped into universality classes similar to equilibrium systems. As we have shown, the optomechanical system undergoes both continuous and discontinuous DPTs in the thermodynamic limit. To specify the universality class we numerically compute the critical exponents of two quantities (see Fig.~\ref{fig:powerlaw1}): photon number fluctuations $\langle\delta a^\dagger\delta a\rangle\propto \abs{G-G_c}^{\nu_{c}}$ and asymptotic decay rate (ADR) $\kappa_{ADR}\propto\abs{G-G_c}^{\nu_{ADR}}$ (with $\nu_c=-1$ and $\nu_{ADR}=1$), where the ADR determines the time scale at which the steady-state is attained~\cite{ADR}. %The numerical results are presented in Fig.~\ref{fig:powerlaw1}.

Finally, we calculate the finite-size scaling of optomechanical systems with the Keldysh formalism~\cite{keldysh1,keldysh3} (see Appendix \ref{keldysh}), which gives rise to the following scaling relation for the photon number in the cavity 
\begin{equation}
\langle n_c\rangle\approx\langle x_{cl}^2\rangle\sim N^{1/3}.
\end{equation}
We find that optomechanical systems have the same critical exponents as Dicke and Rabi models~\cite{Rabi,Dicke}, but a different finite size-scaling. This should not be surprising since all these models have the same type of interaction $\sim(a+a^\dagger)(b+b^\dagger)$ in the exact thermodynamic limit (in the Rabi model the interaction is $\sim(a+a^\dagger)^2$). In contrast, their nonlinear interactions, which are the central element for deriving the finite scaling, are different (quartic versus cubic).

\section{Conclusion and outlook}
 In this paper, we show that driven dissipative optomechanical systems undergo dissipative phase transitions in the well-defined thermodynamic limit. The system exhibits a rich phase diagram composed of different regions which are separated by DPTs: first- and second-order dissipative phase transitions, and periodic orbits. 

Some of the first-order DPTs happen in a regime of  parameters that are within experimental reach. Our work predicts the observation of first-order DPT and its characteristics, such as critical slowing down and hysteresis behavior in the laboratory. Furthermore, our theoretical analysis at the thermodynamic limit shows that the optomechanical system is stable in the ultrastrong coupling regime  (more details in Appendix~\ref{eigs}). This should allow experimental studies to probe the physics related to the ultrastrong coupling, such as the nontrivial ground state~\cite{usc}.
 
Optomechanical systems do exhibit second-order DPTs with and without bifurcation.
The equilibrium counterpart of the second-order DPT without bifurcation cannot be explained by Landau theory, and in general they correspond to the domain of topological phase transitions~\cite{topology}. This is a novel topic in the context of out of equilibrium quantum systems, and our work may help to improve the understanding of the matter.

We conclude with a final remark on the possible relevance of our study for the domain of quantum computing. The fact that the system may exhibit multiple steady states is an alternative way to utilize a harmonic oscillator's infinite-dimensional Hilbert space \cite{Mazyar} together with the symmetry of the optomechanical steady-state for encoding and processing quantum information.
 One can achieve this goal by suitably engineering the initial state for the parameter regimes that undergo second-order DPT with bifurcation.

\textbf{\textit{Acknowledgment}}. We thank Joshua Morris
for helpful comments. B.D. acknowledges support from the Austrian Science Fund (FWF) through BeyondC-F7112.  U.D. and M.A. acknowledge support by the Austrian Science Fund (FWF, Project No. I 5111-N), the European Research Council (ERC 6 CoG QLev4G), by the ERA-NET programme QuantERA under the Grants QuaSeRT and TheBlinQC (via the EC, the Austrian ministries BMDW and BMBWF and research promotion agency FFG), by the European Union’s Horizon 2020 research and innovation programme under Grant No. 863132 (iQLev).

\appendix
\section{Eigenvalues of the drift Matrix}\label{eigs}
\begin{figure}[H]
  \centering
       \subfloat[]{\includegraphics[width=4.0cm]{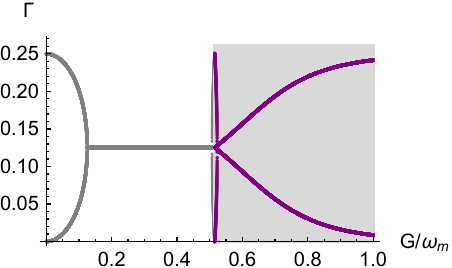}}
     \subfloat[]{\includegraphics[width=4.0cm]{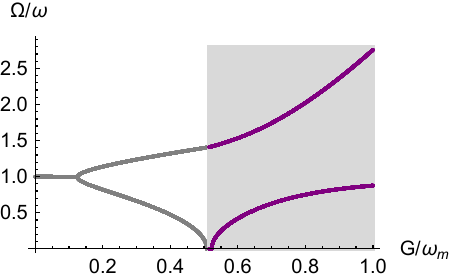}}
    \caption{\textbf{Eigenvalues of the drift matrix.} a) Effective mode dissipation $\Gamma$ in terms of effective coupling $G$. b) Eigenfrequencies $\Omega$ in terms of effective coupling $G$. The gray lines in both plots correspond to the first steady-state solution (I), while the purple ones correspond to the second steady-state solution (II). The gray area indicates the instability region if only the linear coupling is considered~\cite{peterson}. Both plots have been evaluated for $\kappa/\omega_{m}=1/4$, $\gamma/\omega_{m}=0$, and $\tilde{\Delta}/\omega_{m}=-1$ (resonance case).}
    \label{fig:eigenvalues}
\end{figure}

In this section, we study the behavior of the eigenfrequencies and decay rates in the weak, strong, and ultrastrong coupling regime. Since the drift matrix is real, its eigenvalues appear in the complex conjugate pairs. The real part of the eigenvalues corresponds to the effective mode dissipation $\Gamma$. The system is stable only if all dissipation rates are positive (see Fig.~\ref{fig:eigenvalues} (a)). On the other hand, the imaginary part of the eigenvalues of the drift matrix corresponds to the system's eigenfrequencies. In Fig.~\ref{fig:eigenvalues} (b), we plot the eigenfrequencies with respect to the effective coupling $G$, for the resonance case ($\tilde{\Delta}/\omega_m=-1$). By increasing $G$ from $0$, the eigenfrequencies of the combined system split into two separate branches in the strong coupling regime (gray lines). The amount of splitting between these two branches indicates the energy exchange rate, distinguishing between the strong and ultrastrong regimes~\cite{peterson}. As the system coupling approaches the transition point $G^*$, the lower branch goes to zero. In previous studies~\cite{peterson}, the region $G>G^*$ (Fig.~\ref{fig:eigenvalues} gray area) has been considered as the fundamental parametric instability region due to the linear approximation. This assumption introduces the upper bound for the energy exchange rate. In contrast, here, we show that the system is stable in this region. In fact, it is stable with a different steady-state. The purple lines show the eigenfrequencies after the phase transition. %The lower branch converges to $1$ for $G\gg G^*$.
\section{Keldysh Path integral formalism}\label{keldysh}
In this section, we briefly introduce the Keldysh path integral formalism~\cite{keldysh1,keldysh3}. We use this formalism along side with the renormalization group (RG) method~\cite{RG} to determine the finite size scaling exponents.
The most general evolution of the density matrix is given by the master equation
\begin{align}
    \rho(t)=e^{(t-t_0)\mathcal{L}}\rho(t_0)\equiv \lim_{N\rightarrow \infty}(1+\delta_t\mathcal{L})^N\rho(t_0).
\end{align}
Here the Trotter decomposition is used to derive the evolution operator and Keldysh partition function
\begin{align}
    Z=tr[\rho(t)]=1,
\end{align}
which is one of the central objects in this formalism. Within the formalism, the partition function takes the following form~\cite{keldysh1}
\begin{align}
    Z=\int\mathcal{D}[a_+,a_+^*,a_-,a^*_-,b_+,b_+^*,b_-,b^*_-]e^{iS}.
\end{align}
Here $a_\pm$ and $b_\pm$ are complex numbers on the Keldysh contour and $S$ is the Keldysh action. In our case, the action is composed of two parts, free part,
\begin{frame}
\footnotesize
\setlength{\arraycolsep}{0.0000000005pt}
\medmuskip = 0.0000000005mu
\begin{align}
    &S_F=\nonumber\\&a_+^*(i\partial_t+\tilde{\Delta}+i\kappa)a_++a_-^*(-i\partial_t-\tilde{\Delta}+i\kappa)a_--2i\kappa a_+ a_-^*\nonumber\\&+b_+^*(i\partial_t-\omega_m+i\gamma)b_++b_-^*(-i\partial_t+\omega_m+i\gamma)b_--2i\gamma b_+ b_-^*,
\end{align}
\end{frame}
and the interaction part,
\begin{align}
    S_I=&-(\tilde{\alpha}\,a_+^*+\tilde{\alpha}^*\,a_+)(b_++b_+^*)-g\,a_+^* a_+(b_++b_+^*)\nonumber\\&+(\tilde{\alpha}\,a_-^*+\tilde{\alpha}^*\,a_-)(b_-+b_-^*)+g\,a_-^*a_-(b_-+b_-^*).
\end{align}
It is convenient to introduce Keldysh rotation, $a_{cl}=1/\sqrt{2}\,(a_++a_-)$, $a_{q}=1/\sqrt{2}\,(a_+-a_-)$, $b_{cl}=1/\sqrt{2}\,(b_++b_-)$, $b_{q}=1/\sqrt{2}\,(b_+-b_-)$. Now, let us write down the action in the frequency space and the rotated basis
\begin{align}
    S=\frac{1}{2}\int_\omega V^\dagger(\omega)\begin{pmatrix}
0 & [G^A]^{-1}(\omega)\\
[G^R]^{-1}(\omega)& D^K
\end{pmatrix}V(\omega),
\end{align}
where
\begin{align}
    V(\omega)=\begin{pmatrix}
    a_{cl}(\omega)\\a^*_{cl}(-\omega)\\b_{cl}(\omega)\\b^*_{cl}(-\omega)\\
    a_{q}(\omega)\\a^*_{q}(-\omega)\\b_{q}(\omega)\\b^*_{q}(-\omega)\end{pmatrix}.
\end{align}
The Block entry, $G^R$ is the retarded Green's function given by:
\begin{frame}
\footnotesize
\setlength{\arraycolsep}{0.0000000005pt}
\medmuskip = 0.0000000005mu
\begin{align}
    &[G^R]^{-1}(\omega)=\\\nonumber&\begin{pmatrix}\omega-\tilde{\Delta}+i\kappa&0&-\tilde{\alpha}&-\tilde{\alpha}\\0&-\omega-\tilde{\Delta}-i\kappa&-\tilde{\alpha}^*&-\tilde{\alpha}^*\\-\tilde{\alpha}^*&-\tilde{\alpha}&\omega-\omega_m+i\gamma&0\\-\tilde{\alpha}^*&-\tilde{\alpha}&0&-\omega-\omega_m-i\gamma\end{pmatrix},
\end{align}
\end{frame}while $G^A=(G^R)^*$ is the advanced Green's function, and $D^k=2i\,diag(\kappa,\kappa,\gamma,\gamma)$ is the Keldysh Green's function. These Green functions shall serve to derive the properties of the system. Namely, characteristic frequencies of the system, $\omega$ are given by the zeros of the determinant of $[G^R]^{-1}(\omega)$, and they are directly related to the eigenvalues of the drift matrix, i.e. $\lambda=i\omega$. To derive the finite size scaling exponent for optomechanical system far from thermodynamic limit we will proceed with the same approach as for the open Dicke and Rabi models~\cite{Dicke,Rabi}. Firstly, we shall integrate out the mechanics mode and change the variable to the $a_{cl(q)}=\sqrt{-\tilde{\Delta}/2}(x_{cl(q)}+ip_{cl(q)})$ and $a^*_{cl(q)}=\sqrt{-\tilde{\Delta}/2}(x_{cl(q)}-ip_{cl(q)})$, then we integrate out the $p_{cl}$ and $p_{q}$. Finally by low frequency expansion we arrive at the following form for the action in the time domain
\begin{align}
    &S^c_{xx}=\int dt\begin{pmatrix}x_{cl}(t)&x_q(t)\end{pmatrix}\\\nonumber&\begin{pmatrix}0&-2i\kappa \partial_t\\2i\kappa\partial_t&-2i\kappa \tilde{\Delta}(1+\kappa^2/\tilde{\Delta}^2)\end{pmatrix}\begin{pmatrix}x_{cl}(t)\\x_q(t)\end{pmatrix}.
\end{align}
This action is invariant under following scaling transformations
\begin{align}
    t\rightarrow r t, \,\,\,\,x_{cl}(t)\rightarrow\sqrt{r}x_{cl}(t),\,\,\,\,  x_{q}(t)\rightarrow\frac{1}{\sqrt{r}}x_{q}(t).
\end{align}
The same analysis holds for the mechanics mode. Now if we look at the nonlinear interaction up to $g=G/\abs{\alpha}=G/\sqrt{N}$ order, the latter scaling transformation yields to the following nonlinear terms of the action
\begin{align}\label{clv}
&\frac{G}{\sqrt{N}}\int dt\,\phi_{cl}\phi_{cl}\phi_q\rightarrow \frac{G r\sqrt{r}}{\sqrt{N}}\,\int dt\,  \phi_{cl}\phi_{cl}\phi_q,\\\label{qv}
&\frac{G}{\sqrt{N}}\int dt\,\phi_{q}\phi_q\phi_q\rightarrow \frac{G}{\sqrt{r}\sqrt{N}} \int dt\,  \phi_{q}\phi_q\phi_q.
\end{align}
Here $\phi$ represent any of the $a$, $a^*$, $b$, and $b^*$ fields. The scaling above shows that the classical vertex Eq. \ref{clv} is relevant and the quantum vertex Eq. \ref{qv} is irrelevant in the renormalization group (RG) flow. Therefore, quantum vertex contribution is negligible compared to the classical vertex contribution in the limit $N\gg 1$. Now, in order to make the theory scale invariant at the continuous transition point we need to renormalize $N$ in the following way
\begin{align}
    N\rightarrow N^\prime=\frac{N}{r^3}.
\end{align}
This results in the following scaling relation for the photon number in the cavity $\langle n_c\rangle\approx\langle x_{cl}^2\rangle\sim N^{1/3}$.

\end{document}